\documentclass[3p,times]{elsarticle}
\usepackage{ecrc}
\usepackage{graphicx}
\usepackage{epsfig}

\volume{00}
\firstpage{1}
\journalname{Nuclear Physics A}
\runauth{Lusaka Bhattacharya et al.}
\jid{nupha}

\usepackage{amssymb}
\usepackage{amsthm}
\usepackage{lineno}

\biboptions{comma,round}
\biboptions{}

\usepackage[figuresright]{rotating}
\begin{document}
\begin{frontmatter}
\dochead{}

\title{Estimation of isotropization time ($\tau_{\rm iso}$) of QGP from direct
  photons}

\author{Lusaka Bhattacharya}

\address{Saha Institute of Nuclear Physics\\ 
1/AF Bidhannagar, Kolkata, Pin
  700064, India}

\begin{abstract}
We calculate transverse momentum distribution of direct photons 
from various sources by taking into account the initial state 
momentum anisotropy of quark gluon plasma (QGP). 
The total photon yield is then compared with the recent measurement of 
photon transverse momentum distribution by the PHENIX collaboration. 
It is also demonstrated that the presence of such an anisotropy 
can describe the PHENIX photon data better than the isotropic case 
in the present model. We show that the isotropization time thus extracted 
lies within the range $1.5 \geq\tau_{\rm iso} \geq 0.5$ fm/c for the 
initial condition used here.
\end{abstract}
\end{frontmatter}

\section{Introduction }

The primary goal of relativistic heavy ion collisions is to create
a new state of matter, called quark gluon plasma  and to study
its properties through various indirect probes. Out of all the
properties of the QGP, the most difficult problem lies in the
determination of isotropization and thermalization time scales
($\tau_{\rm iso}$ and $\tau_{\rm therm}$). Studies on elliptic 
flow (upto about $p_T \sim 1.5 - 2 $ GeV) using ideal hydrodynamics 
indicate that the matter produced in such collisions becomes isotropic 
with $\tau_{\rm iso} \sim 0.6$ fm/c~\cite{Pasi}. On the 
other hand, using second order transport coefficients with conformal 
symmetry it is found that the isotropization/thermalization time has 
sizable uncertainties~\cite{0805.4552_ref4}. Consequently, there are 
uncertainties in the initial temperature as well. Electromagnetic 
probes have been proposed to be one of the most promising tools to 
characterize the initial state of the collisions~\cite{ann,jpr}. Because 
of the very nature of their interactions with the constituents of the 
system they tend to leave the system without much change of their 
energy and momentum. In fact, photons (dilepton as well) can be 
used to determine the initial temperature, or equivalently the 
equilibration time.

It is to be noted that while estimating photons from 
QGP~\cite{janejpg,dksepjc,turbide_gale_fro_heinz}, it is assumed 
that the matter formed in the relativistic heavy ion collisions 
is in thermal equilibrium. The measurement of elliptic flow parameter 
and its theoretical explanation also support this assumption. On the 
contrary, perturbative estimation suggests the slower thermalization 
of QGP~\cite{PRC75_ref2}. However, recent hydrodynamical 
studies~\cite{0805.4552_ref4} have shown that due to the poor 
knowledge of the initial conditions there is a sizable amount of 
uncertainty in the estimate of thermalization or isotropization time. 
In view of the 
absence of a theoretical proof behind the rapid thermalization and the 
uncertainties in the hydrodynamical fits of experimental data, such an 
assumption may not be justified. Hence in stead of equating the 
thermalization/isotropization time to the QGP formation time, in this 
work, we will introduce an intermediate time scale 
(isotropization time, $\tau_{\rm iso}$) to study the effects of early 
time momentum-space anisotropy on the total photon yield and compare 
it with the PHENIX photon data~\cite{adler1,adler2,phenix08}.
Recently, it has been shown in 
Ref.~\cite{mauricio} that for fixed initial conditions, the introduction 
of a pre-equilibrium momentum-space anisotropy enhances high energy 
dileptons by an order of magnitude.  In case of photon transverse momentum
distribution similar results have been reported for
various evolution scenarios~\cite{lusakaprc}.

The plan of the paper is the following. In the next section
we will discuss the mechanisms of
photon production from various possible sources and
the space-time evolution of the matter very briefly.
Section 3 is devoted  to describe the results for
various initial conditions and we summarize in section 4.

\section{Formalism}

\subsection{Photon rate: Anisotropic QGP}

The lowest order processes for photon emission from QGP are the
Compton ($q ({\bar q})\,g\,\rightarrow\,q ({\bar q})\,
\gamma$) and the annihilation ($q\,{\bar q}\,\rightarrow\,g\,\gamma$)
processes. 
The rate of photon production from 
anisotropic plasma due to these processes has been 
calculated in Ref.~\cite{prd762007}. The soft contribution is calculated by
evaluating the photon polarization tensor for an oblate momentum-space
anisotropy of the system where the cut-off scale is fixed at 
$k_c \sim  \sqrt g p_{hard}$. Here $p_{hard}$ is a hard-momentum scale 
that appears in the distribution functions. 
The differential photon production rate for $1+2\to3+\gamma$ processes in 
an anisotropic medium is given by~\cite{prd762007}: 
\begin{eqnarray} 
E\frac{dN}{d^4xd^3p}&=& 
\frac{{\mathcal{N}}}{2(2\pi)^3} 
\int \frac{d^3p_1}{2E_1(2\pi)^3}\frac{d^3p_2}{2E_2(2\pi)^3}
\frac{d^3p_3}{2E_3(2\pi)^3}
f_1({\bf{p_1}},p_{\rm hard},\xi)f_2({\bf{p_2}},p_{\rm hard},\xi) \nonumber\\
&\times&(2\pi)^4\delta(p_1+p_2-p_3-p)|{\mathcal{M}}|^2 
[1\pm f_3({\bf{p_3}},p_{\rm hard},\xi)]
\label{photonrate}
\end{eqnarray}
where, $|{\mathcal{M}}|^2$ represents the spin averaged matrix element
squared for one of those processes which contributes to the photon rate
and ${{\mathcal N}}$ is the degeneracy factor of the corresponding
process. $\xi$ is a parameter controlling the strength of the anisotropy 
with $\xi > -1$. $f_1$, $f_2$ and $f_3$ are the anisotropic 
distribution functions of the medium partons. 
Here it is assumed that the infrared 
singularities can be shielded by the thermal masses for the 
participating partons. This is a good approximation at short times  
compared to the time scale when plasma instabilities start to play 
an important role. 
The anisotropic distribution function can be obtained~\cite{stricland} 
by squeezing or stretching an arbitrary isotropic distribution function 
along the preferred direction in momentum space,
%
%\begin{eqnarray}
$f_{i}({\bf k},\xi, p_{hard})=f_{i}^{iso}(\sqrt{{\bf k}^{2}+\xi 
({\bf k.n})^{2}},p_{hard}),
$
%\label{dist_an}
%\end{eqnarray}
%
where ${\bf n}$ is the direction of anisotropy. It is important to
notice that $\xi > 0$ corresponds to a contraction of the 
distribution function in the direction of 
anisotropy and $-1 < \xi < 0 $ corresponds to a stretching in the
direction of anisotropy. In the context of relativistic
heavy ion collisions, one can identify the direction of anisotropy with 
the beam axis along which the system expands initially. The hard momentum
scale $p_{hard}$ is directly related to the average momentum of the 
partons. In the case of an isotropic QGP, $p_{hard}$ can be identified 
with the plasma temperature ($T$). 

\subsection{Photon rate : Isotropic case} 

As mentioned earlier the QGP evolves  hydrodynamically from 
$\tau_{\rm iso}$ onwards. In such case the distribution functions become 
Fermi-Dirac or Bose-Einstein distributions. The photon emission rate, 
in isotropic case, from Compton 
($q ({\bar q})\,g\,\rightarrow\,q ({\bar q})\,\gamma$) and annihilation 
($q\,{\bar q}\,\rightarrow\,g\,\gamma$) processes has been 
calculated from the imaginary part of the photon self-energy by 
Kapusta et al.~\cite{kap} in the 1-loop approximation. However, it has 
been shown by Auranche et al.~\cite{aur1} that the two loop contribution 
is of the same order as the one loop due to the shielding of infra-red 
singularities. The complete calculation upto two loop was done by 
Arnold et al.~\cite{arnold}. 
In this paper we have calculated the photon production rate from hot hadronic
matter. We follow the calculations done in Ref.~\cite{turbide}
where  convenient parameterizations have been given for the reactions
considered. These parameterizations will be used while doing the space-time
evolution to calculate the photon yield from meson-meson reactions.
The photon emission rate (static) from reactions of the type 
$B\,M\,\rightarrow\, B\,\gamma$ ($B$ denotes baryon) has been 
calculated in Ref.~\cite{rap2}. It is 
shown that this contribution is not negligible compared to that
meson-meson reactions. To evaluate 
photon rate due to nucleon (and antinucleon) scattering from 
$\pi$, $\rho$, $\omega$, $\eta$ and $a_1$ mesons in the thermal bath we 
use the phenomenological interactions described in Ref.~\cite{rap2}. 
Besides the thermal photons from QGP and hadronic matter we also
calculate photons from initial hard scattering from the reaction of the 
type $h_A\,h_B\,\rightarrow\,\gamma\,X$ using perturbative QCD. We include
the transverse momentum broadening in the initial state 
partons~\cite{wong,owens}. 

\subsection{Space-time evolution} 

The expected total photon rate must be convoluted with the space-time 
evolution of the fireball. The system evolves anisotropically from 
$\tau_i$ to $\tau_{\rm iso}$ where one needs to know the time 
dependence of $p_{\rm hard}$ and $\xi$. We have used a phenomenological 
model~\cite{mauricio,lusakaprc} to describe the time dependence of 
$p_{hard}$ and $\xi$. In the frame work of 
this model, $\xi=0$ at $\tau=\tau_i$ and it grows with time ($\tau$) and 
reaches maximum at $\tau=\tau_{iso}$, after that $\xi$ decreases to 
zero at $\tau>>\tau_{iso}$.   
We shall follow the work of Ref.~\cite{mauricio,lusakaprc} to 
evaluate the $p_T$ distribution of 
photons from the first few Fermi of the plasma evolution. In our 
calculation, we assume a first-order phase transition beginning at the 
time $\tau_c (p_{\rm hard}(\tau_c)=T_c)$ and ending at $\tau_{H}=r_d\tau_c$ 
where $r_d=g_Q/g_H$ is the ratio of the degrees of freedom in the two 
(QGP phase and hadronic phase) phases. Therefore, the total 
thermal photon yield, arising from the present scenario is given by,
\begin{equation}
\frac{dN}{d^2p_Tdy}=
\left[\int\,d^4x\, E\frac{dR}{d^3p}\right]_{\rm aniso} + 
\left[\int\,d^4x\, E\frac{dR}{d^3p}\right]_{\rm hydro}, 
\label{yield_total} 
\end{equation}
where the first term denotes the contribution from the anisotropic
QGP phase and the second term represents the contributions 
evaluated in ideal hydrodynamics scenario.

\section{Results}

\begin{figure}
\centering
\includegraphics[height=5.7cm]{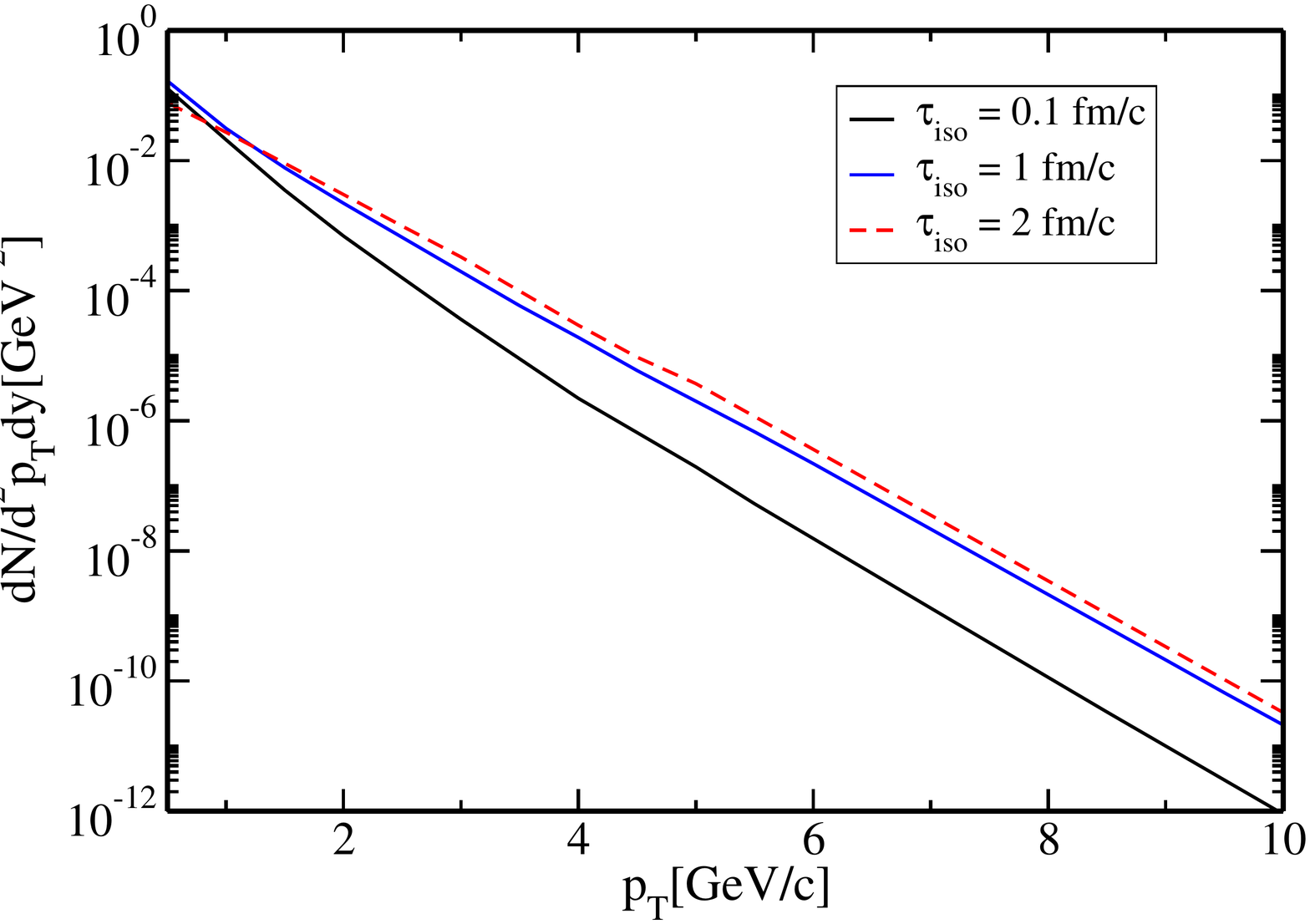}
\caption{(Color online) Medium photon spectrum, $dN/d^2p_Tdy$, at $y=0$ 
for the {\em free-streaming interpolating} model ($\delta = 2$) 
for three different values of isotropization time, $\tau_{iso}$, with 
initial conditions,$T_i= 440$ Gev and $\tau_{\i}=0.1$ fm/c.}
\label{fig1}      
\end{figure}

We have considered the initial condition, $T_i=440$ MeV, $\tau_i=0.1$ fm/c 
and {\em free-streaming interpolating} model 
($\delta =2 $)~\cite{lusakaprc, lusakaprc_2} for the 
pre-equilibrium evolution. In this initial condition the maximum value of 
$\xi$ will be $\sim 70$ at $\tau=\tau_{iso}$. 
In Fig.~(\ref{fig1}) we present the
photon yield due to Compton and annihilation processes in the mid 
rapidity ($\theta_{\gamma}=\pi/2$, $\theta_{\gamma}$ being the angle
between the photon momentum and the anisotropy direction) 
as a function of photon transverse momentum. 
In estimating this result, we have used
$\alpha_s=0.3$. Different lines in Fig.~\ref{fig1} correspond to 
different isotropization times, $\tau_{\rm iso}$. We clearly
observe enhancement of photon yield when $\tau_{\rm iso}>\tau_i$. The 
enhancement of photon yield in the transverse directions 
($y = 0$) is due to the fact that momentum-space anisotropy 
enhances the density of plasma partons moving at the mid 
rapidity~\cite{lusakaprc}.
\begin{figure}
\centering
\includegraphics[height=5.7cm]{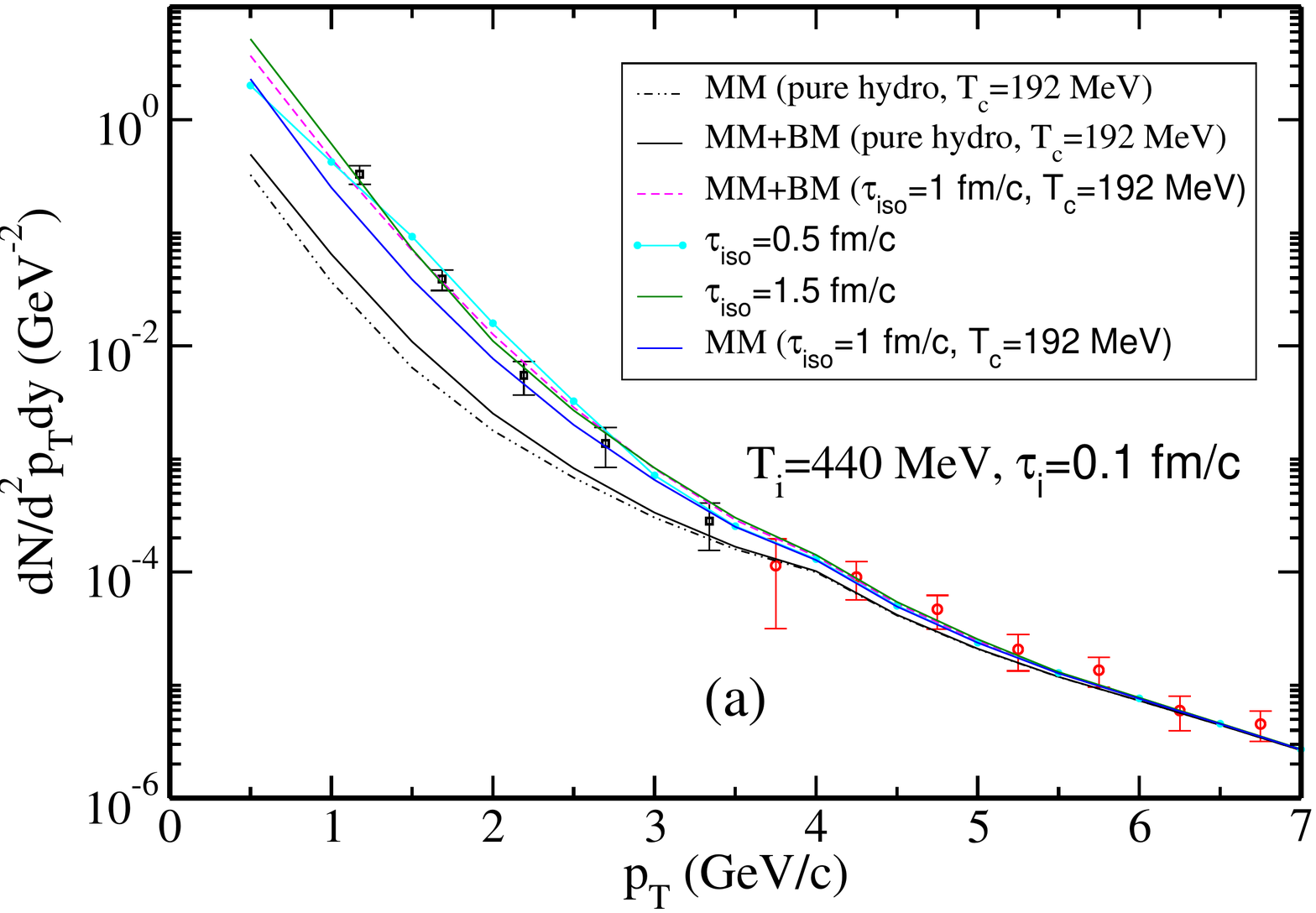}
\includegraphics[height=5.7cm]{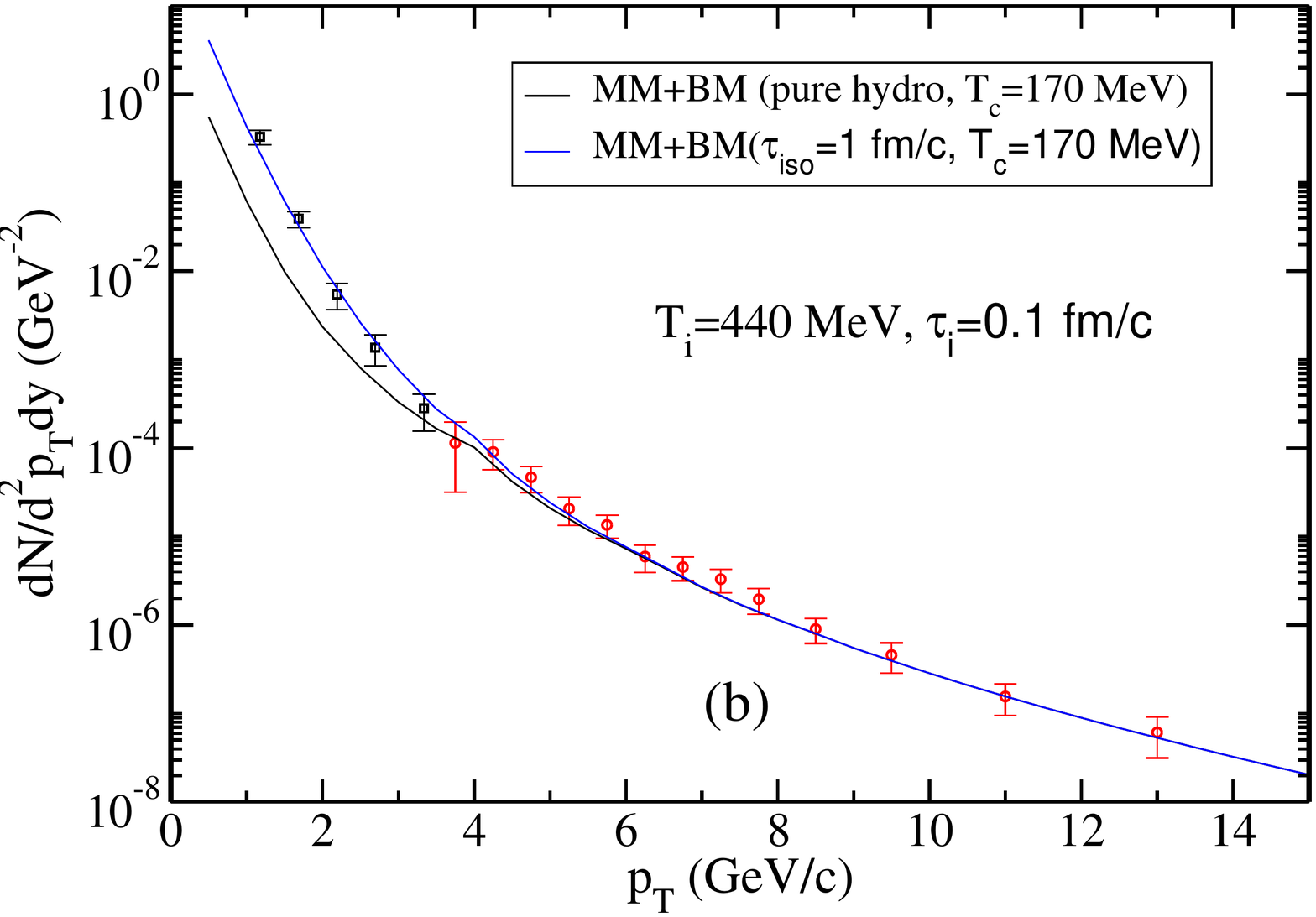}
\caption{(Color online) Photon $p_T$ distributions at 
RHIC energies with initial condition $T_i=440$ GeV, for 
(a) $T_c$ = 192 MeV and (b) 170 MeV.}
\label{fig3}      
\end{figure}
To show that the presence of initial state momentum anisotropy and the
importance of the contribution from baryon-meson reactions we plot the
the total photon yield assuming hydrodynamic evolution from the very 
begining as well
as with finite $\tau_{\rm iso}$ (right panel describes the total 
contribution with and without
the initial state momentum space anisotropy only for
$\tau_{\rm iso}$ = 1 fm/c) in Fig.~(\ref{fig3}). 
It is clearly seen that some amount of anisotropy is needed to reproduce 
the data. We note that the value
of $\tau_{\rm iso}$ needed to describe the data also lies in the range
1.5 fm/c$ \geq\tau_{\rm iso} \geq 0.5$ fm/c for both values of the transition
temperatures.  

\section{Conclusion}

To summarize, we have calculated total single photon transverse momentum
distributions by taking into account the effects of the pre-equilibrium
momentum space anisotropy of the QGP and late stage transverse expansion
on photons from hadronic matter with various initial conditions.  
To describe space-time evolution in the very early stage
we have used the phenomenological model described in Ref.~\cite{mauricio} 
for the time
dependence of the hard momentum scale ($p_{hard}$) and plasma 
anisotropy parameter ($\xi$). To calculate the hard photon contributions
we include the transverse momentum broadening in the initial hard scattering. 
The total photon yield is then compared with the PHENIX photon data. 
Within the ambit of the present model it is shown that the data can be 
described
quite well if $\tau_{\rm iso}$ is in the range of 0.5 - 1.5 fm/c for all 
the combinations of initial conditions and transition temperatures 
considered here. 
It is to be noted that the apparent hump observed in all the figures 
(except Fig.(5)) needs to be understood and we wish to discuss 
it in a subsequent paper. 

\label{}

\bibliographystyle{elsarticle-num}
\bibliography{<your-bib-database>}

\end{document}